\documentclass[aps,prl,twocolumn,superscriptaddress,showpacs]{revtex4} 
\usepackage[dvips]{graphics}
\usepackage[german,english]{babel} 
\usepackage{amssymb} 
\usepackage{amsfonts} 
\newcommand{\eqa}{\begin{eqnarray}}
\newcommand{\eqe}{\end{eqnarray}}
\newcommand{\la}{\langle}
\newcommand{\ra}{\rangle}

\newcommand{\B}{\beta}
\newcommand{\C}{\gamma}
\newcommand{\D}{\delta}
\newcommand{\DD}{\Delta}

\begin{document} 
 
\title{Probing molecular free energy landscapes by periodic loading}

\author{Oliver Braun} 
\affiliation{II. Institut f\"ur Theoretische Physik, Universit\"at 
Stuttgart, 70550 Stuttgart, Germany}
\author{Andreas Hanke} 
\affiliation{II. Institut f\"ur Theoretische Physik, Universit\"at 
Stuttgart, 70550 Stuttgart, Germany}
\affiliation{Theoretical Physics, University of 
Oxford, 1 Keble Road, Oxford OX1 3NP, UK}
\author{Udo Seifert} 
\affiliation{II. Institut f\"ur Theoretische Physik, Universit\"at 
Stuttgart, 70550 Stuttgart, Germany}
 
 
\begin{abstract} 
Single molecule pulling experiments provide information about 
interactions in biomolecules that cannot be obtained by any other 
method. However, the reconstruction of the molecule's free energy 
profile from the experimental data is still a challenge, in 
particular for the unstable barrier regions. We propose a new
method for obtaining the full profile by introducing a 
periodic ramp and using Jarzynski's identity for obtaining 
equilibrium quantities from non-equilibrium data. Our simulated 
experiments show that this method delivers significant more 
accurate data than previous methods, under the constraint of 
equal experimental effort.
\end{abstract} 

 
\maketitle 

\vspace*{-7.5mm}

{\sl Introduction -} A key feature of biological systems is the high degree of 
self-organization 
of polymers, proteins, and other macromolecules, and their interaction with
smaller components such as energy providers or messenger molecules \cite{alberts}. 
These processes are ultimately driven by specific and tunable molecular 
interactions. Their detailed knowledge is thus a prerequisite for the 
understanding how biological systems work on molecular and higher levels.
Recent developments of highly sensitive force probes such as atomic force 
microscopy (AFM) \cite{FMG94,CM02}, 
optical and magnetic tweezers \cite{ot,SCB00,Dan03}, and biomembrane 
force probes \cite{bmp,Mer99} make it possible to probe the molecular 
interactions of individual biomolecules by their response to mechanical 
stress (see \cite{Merk01,RG02} for reviews). The systems 
studied by single molecule pulling experiments can be 
divided in two groups: in {\em rupture} experiments, receptor and ligand
molecules are attached to a substrate and a transducer, respectively,
often via chemical linkers. After allowing receptors and ligands to 
bind, the transducer, e.g., an AFM cantilever, is pulled away, which 
causes the receptor-ligand pairs to rupture. The maximum
force the molecule can withstand has been measured
in this way for biotin and streptavidin \cite{FMG94,Mer99}, and many 
other receptor-ligand pairs \cite{Wei03}.
Secondly, {\em un- and refolding} experiments probe the elastic properties 
of an individual biomolecule. The molecule is attached between a 
substrate and a transducer, again via chemical linkers. 
Force-extension relations are obtained by measuring the force as a 
function of the position of the transducer. 
In this way one may explore regions of the free energy landscape of the
biomolecule far away from thermal folding pathways.
Investigated systems include DNA \cite{MS95,stru,wil}, 
RNA \cite{Lip01,Har03}, polysaccharides \cite{ROH97},
the muscle protein titin \cite{rief97},
the membrane protein bacteriorhodopsin (BR) \cite{mueller}, 
and many other proteins \cite{ZR03}.

\begin{figure}[h]
{\par\centering \scalebox{.75}
{\includegraphics{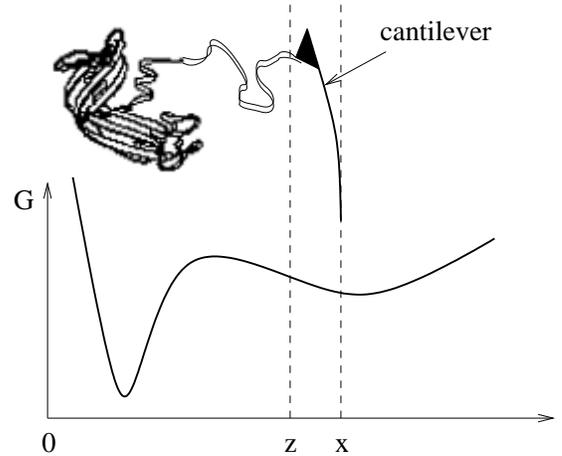}} 
\par}
\caption{\label{bez9.fig} Schematic view of the experimental setup 
and a generic free energy potential $G(z)$. The first minimum 
represents the folded state, whereas the second shallow minimum 
represents the unfolded state of the biopolymer. The coordinate 
$x$ denotes the position of the cantilever and $z$ the position 
of the cantilever tip to which one end of the biopolymer 
is attached.}  
\end{figure}

Figure \ref{bez9.fig} shows a typical setup of single molecule pulling 
experiments. The molecule is attached between the substrate surface and
the cantilever tip. The position of the cantilever $x(t)$ is moved according 
to a prescribed experimental protocol. The extension of the molecule is 
described by a suitable reaction coordinate \, $z$ \, given by

\noindent
the position of the cantilever tip. For fixed extension $z$
and time $t$, the energy
of the molecule perturbed by the cantilever spring is given by the 
time-dependent Hamiltonian
\begin{equation} \label{ham}
H(z,t) = G(z) + V_0(z,x(t))\equiv G(z) + 
\frac{k}{2} \left( x(t) - z \right)^2 \, \, ,
\end{equation}
where $G(z)$ is the free energy profile of the unperturbed molecule. 
The second term describes the external force acting on the molecule 
in terms of a harmonic potential with effective spring constant $k$. 
Since the molecule is coupled to a heat bath at temperature $T$ the 
time evolution of $z$ is stochastic.

Traditionally, the cantilever is moved according to a linear ramp,
\begin{equation} \label{linear}
x(t) = x_0 + v t \, \, ,
\end{equation}
with offset $x_0$ at $t = 0$ and constant velocity $v$, and the force 
$F(t,v)$ acting on the cantilever is recorded.
The challenge is to recover from these data the unperturbed molecule's 
free energy profile $G(z)$ containing the desired information about the 
molecular interactions. Evans and Ritchie first pointed out that the 
rupture force of receptor-ligand pairs depends on the loading rate $v$ 
\cite{ER97}. Thus, by combining the data $F(t,v)$ for a broad spectrum 
of loading rates $v$, referred to as dynamic force spectroscopy, important 
features of $G(z)$ can be determined such as the distance between the 
minimum and maximum of an energy barrier for rupture \cite{ER97}. 
Heymann and Grubm\"uller refined this technique and obtained the heights 
and positions of the maxima of a molecular force profile $\partial_z G(z)$ 
with high spatial resolution \cite{HG00}. On the other hand, traditional 
experimental protocols like the linear ramp (\ref{linear}) still entail
certain drawbacks. First of all, the thermodynamically 
unstable (concave) barrier regions of $G(z)$, determined by specific 
molecular interactions and therefore of particular interest, are poorly 
sampled due to snapping motion and thus hard to determine \cite{HG00}. 

\begin{figure}[h]
{\par\centering \scalebox{.7}
 {\includegraphics{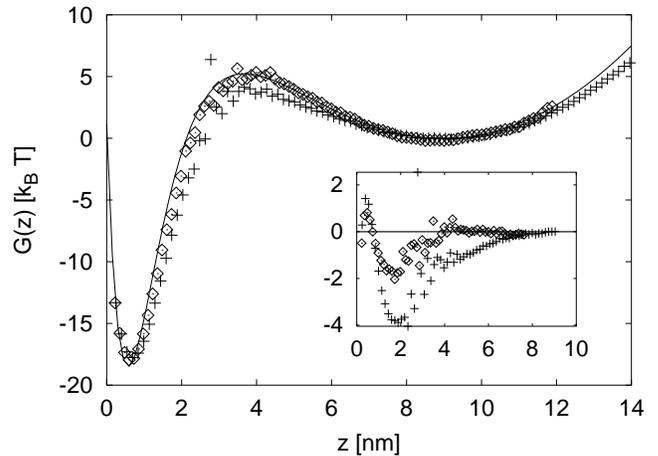}} 
\par}
\caption{\label{1201_3a.fig} 
Comparison of reconstructed free energy profiles $G(z)$ by using 
periodic ($\Diamond$) vs linear loading ($+$), generated by (\ref{jarz1}).
The solid line is the 
original free energy profile $G(z)$. For both methods, 10 trajectories 
of $75$ ms length, a spring constant $k=11.6$ pN/nm, and a 
diffusion constant $D=10^{-7} \mbox{cm}^2/$s were used
[see (\ref{ham}) and (\ref{langevin})]. For periodic 
loading we used the optimal frequency $\omega^\ast= 1.2 \times 10^3$ 1/s
(determined in Fig.\,\ref{opti.fig}), the preloading offset $x_0=6$ nm,
and the amplitude $a=5$ nm in (\ref{cl}).
For linear loading we used the optimal velocity $v^\ast=200$ nm/s
(determined in Fig.\,\ref{opti.fig}). 
The inset compares the relative deviation of the reconstructed data 
from the  original profile $G(z)$ corresponding to the zero line.}
\end{figure}

{\sl Periodic loading -} In an effort to improve the quality of data
obtained by single molecule pulling experiments, in this work we 
propose a new method for obtaining the full free energy profile $G(z)$  
by introducing a {\em periodic} ramp,
\begin{equation} \label{cl}
x(t) = x_0 + a \sin(\omega t) \, \, ,
\end{equation}
with given offset $x_0$, amplitude $a$, and frequency $\omega$. 
Figure \ref{1201_3a.fig} shows that periodic loading delivers  
significant more accurate data for the sample free energy profile
$G(z)$ than the linear ramp
(\ref{linear}), under the constraint of equal experimental effort. 
The improvement of the quality of data in the important barrier 
region of $G(z)$ is striking. 
The better performance of the periodic loading method as compared 
to linear loading  is mainly due to the fact that
 periodic loading ensures that the barrier region is 
traversed often and from both sides. The quality of our reconstruction
moreover depends crucially on the fact that we sample the barrier
region under non-equilibrium conditions taking advantage of
Jarzynski's relation to recover the equilibrium profile \cite{Jar,HS01}.
Driving the system out of equilibrium is important 
 since under quasi-static conditions
 an efficient sampling of the barrier 
region (where $G(z) \gg k_B T$) is inhibited by the 
equilibrium Boltzmann factor $\exp[- G(z)/(k_B T)] \ll 1$.
For periodic loading, the 
freedom to choose the frequency $\omega$ large enough allows one 
to probe the same region under non-equilibrium conditions, thus
overriding the exponential punishment by the equilibrium Boltzmann 
factor.
 The optimal frequency  arises from balancing competing effects
as quantified in a case study
below. This frequency should not be too large in order to
enable the system to follow the external drive. Moreover, 
 the Jarzynski procedure converges the slower the further one is 
away from equilibrium \cite{zuc}. For too small $\omega$,
on the other hand, one does
not generate enough crossings under the constraint of a 
finite total measuring time.

{\sl Simulated experiments -}
We have tested  our proposal of periodic loading with simulated 
experiments and compared it with the traditional method of linear loading.
To this end we have chosen a generic free energy profile $G(z)$ for 
the unfolding of tertiary structures of biopolymers such as the membrane
protein BR \cite{mueller}. Our sample free energy profile has two separated 
minima, one of which is narrow and deep representing the folded state and 
one of which is shallow representing the unfolded state, 
see Fig.\,\ref{bez9.fig}. For BR, a rich structure of unfolding 
transitions under force was found \cite{mueller}.
Single force peaks in the unfolding spectra could be allocated to 
specific changes in molecular configuration. 
Most of the force peaks scatter between $25$ up to $100$ pN. 
With a typical length scale of several nanometers this yields 
an energy barrier of about 20 $k_B T$ at room temperature. 
We focus on one of such transitions and choose a barrier of $2$ nm 
length and 23 $k_B T$ height, leading to a typical transition 
force of about $50$ pN in our simulated pulling experiments. 

For comparing the periodic with the linear ramp, we simulated 
both kinds of protocols using typical parameters as given in the 
caption of Fig.\,\ref{1201_3a.fig}. We have generated an ensemble
of trajectories $\zeta(t)$ of the reaction coordinate $z$ by 
discretizing the Langevin equation
\begin{equation} \label{langevin}
dz/dt= - \gamma^{-1} {\partial} H(z,t)/{\partial z} +\sqrt{2 D}\,\xi(t)
\end{equation}
with the Hamiltonian $H(z,t)$ from (\ref{ham}). The Gaussian
random force $\xi(t)$ has zero mean and short-ranged temporal 
correlations $\la \xi(t) \xi(t') \ra =\D (t-t')$. 
The diffusion constant $D$ is related to the friction coefficient 
$\C$ by the Einstein relation $D=k_B T/\C$. In our simulations,
the length of a time step is limited by the condition that a spatial 
step should be small compared to the typical length scale set by the
free energy profile. In addition, the recording rate of the data 
$\zeta(t)$ should be much smaller than an inverse time step but large 
enough to resolve the cantilever motion.

The reconstructed free energy profile for both protocols is shown 
in Fig.\,\ref{1201_3a.fig}. The overall quality of the data obtained
by the periodic ramp is far better than the linear ramp, especially
in the barrier region where the data obtained by the linear ramp
underestimates the barrier height by several $k_BT$.
 In order to ensure an unbiased comparison, we have chosen the same
total number of trajectories and the same total measuring time 
 for both methods.
In practice, the measuring time of the trajectories, once 
prepared, is not a limiting factor. The periodic ramp therefore allows
measuring as many transitions as necessary to collect the sufficient 
amount of data.

{\sl Reconstruction using Jarzynski's relation -}
Pulling protocols both in real experiments and in simulations used 
here typically generate non-equilibrium data from which one has to 
recover an equilibrium property like $G(z)$.
Furthermore, as outlined above, the method proposed here
purposely takes advantage of the non-equilibrium conditions generated 
by a large enough optimal driving frequency $\omega^{\ast}$, to be
determined below. The difficulty to recover equilibrium properties 
from non-equilibrium data may be resolved by using a recent advance 
in non-equilibrium statistical mechanics due to Jarzynski \cite{Jar}, 
according to which the equilibrium profile $G(z)$ can be inferred by 
suitably averaging non-equilibrium trajectories $\zeta(t)$ of the 
reaction coordinate $z$ \cite{HS01}. This method was verified by 
stretching RNA reversibly and irreversibly between two conformations 
indeed \cite{Lip02}. 

Jarzynski's relation, in general, states that the free energy difference 
$\DD G$ between two equilibrium states can be extracted from averaging 
the work $W$ required to drive the system from one state to the other
according to 
$e^{-\B \DD G}=\la e^{-\B W} \ra$
%
with $\B=1/k_B T$ the inverse temperature \cite{Jar}. 
The generalization  from two states to a $z$-resolved free energy 
profile $G(z)$ perturbed by a harmonic spring (\ref{ham}) reads \cite{HS01}
\begin{equation} \label{profile}
e^{- \beta [H(z,t) - G_0]}
= \left\langle \delta[z - \zeta(t)] \,
e^{- \beta W(t)} \right\rangle \, \, .
\end{equation}
The average $\langle ... \rangle$ is over  
infinitely many 
realizations $\zeta(t')$, $0 < t' < t$, of the stochastic trajectory 
of the biopolymer's end position, starting in equilibrium at $t' = 0$ 
and ending at the given position $z$ at $t' = t$, as enforced by the
delta function. The external work is a functional of $\zeta(t')$ and
given by
$
W(t) = \int_0^{t} dt' \,
\partial_{\tau} H(\zeta(t'), \tau)|_{\tau=t'} .$
The constant
$G_0 = - k_B T \ln 
(\int_0^{\infty} dz' \, e^{-\beta H(z',0)} /\lambda_T)$
is the free energy in the initial state at $t=0$, where the thermal 
wavelength $\lambda_T= h / \sqrt{2 \pi m k_B T}$ serves for normalization.

Summing up the normalized distributions obtained from (\ref{profile})
at each time slice 
with the method of weighted histograms \cite{FS89} yields the reconstruction 
formula for the unperturbed free energy profile of the molecule 
\vspace*{-7mm}

\begin{widetext}
\eqa
\label{jarz1}
G(z)=-\B^{-1}\ln 
{\sum_t\frac{\la \D (z-\zeta(t))\exp(-\B W(t))\ra}{\la \exp (-\B W(t)) \ra}}
\Big/{\sum_t \frac{\exp[-\B V_0(z,x(t))]}{\la \exp (-\B W(t)) \ra}} \, \, .
\eqe
\end{widetext}

\vspace*{-7mm}
\noindent
Using this expression, we have generated the data shown in 
Fig.\,\ref{1201_3a.fig}.

{\sl Optimization with respect to frequency/velocity -}
The quality of the reconstructed free energy profile depends crucially
on the driving frequency $\omega$ for the periodic ramp and the
velocity $v$ for the linear ramp, respectively. To quantify this 
observation, we calculate the mean square error 
$\sigma^2\equiv \la[{\widetilde G(z)}-G(z)]^2 \ra_{z_i}$ where the 
average is taken over discrete values $z_i$ in the $z$-interval 
under consideration. For clarity, we denote by $\widetilde G(z)$ 
the reconstructed free energy profile based on (\ref{jarz1}).
For a better reconstruction quality, $\sigma^2$  is smaller.
The $z$-interval is chosen from the first to the second minimum.
The total measuring time and the number of trajectories are the
same as before.

\begin{figure}[h]
{\par\centering \scalebox{.7}
 {\includegraphics{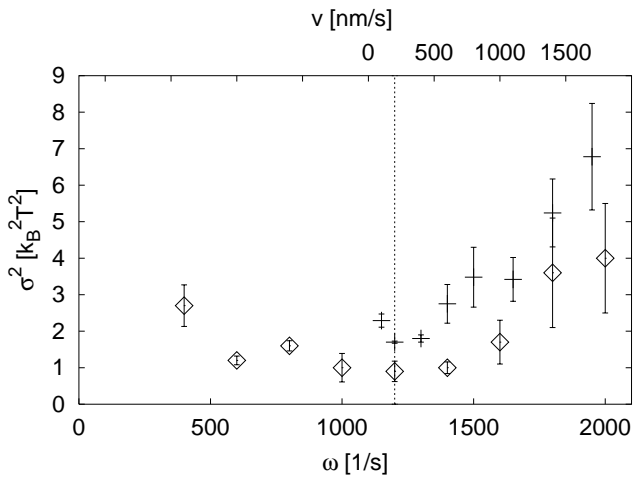}} 
\par}
\caption{\label{opti.fig} 
Mean square error $\sigma^2$ and error bars for reconstructed free 
energy profiles by using the linear ramp (\ref{linear}) as a function 
of $v$ (top scale, $+$), and the periodic ramp (\ref{cl}) as a function of 
$\omega$ (bottom scale, $\Diamond$) (compare Fig.\,\ref{1201_3a.fig}). Both methods 
use 10 runs with 10 trajectories each. The optimal velocity 
$v^\ast$ and frequency $\omega^\ast$, respectively, 
indicated by the vertical dashed line, correspond to the smallest error.}
\end{figure}

Figure \ref{opti.fig} shows the mean square error $\sigma^2$ of the 
reconstructed free energy and some characteristic confidence intervals
(error bars) for both protocols.
For periodic loading, the best results were obtained for the optimal 
driving frequency $\omega^\ast \simeq 1.2 \times 10^3$ 1/s, 
which yields an error of $\sigma^2 \simeq 0.9 \, ({ k_BT})^2$.
This frequency is somewhat smaller than the spontaneous transition rate 
under preloading, which is about $7 \times 10^3$ 1/s for our model system.
By analyzing the work distribution we have convinced ourselves that  
$\omega^\ast$ indeed corresponds to non-equilibrium conditions.
For both smaller and larger frequencies than $\omega^\ast$ the quality of the 
reconstructed data becomes worse as expected from our
reasoning above.

For linear loading, the error increases for increasing driving velocity 
$v$, as expected.
Since we fixed the total measuring time and the total number of 
trajectories, the least 
possible velocity for \, overcoming \, the \, barrier \, is 

\noindent
$v\simeq 100$ nm/s. The smallest error of 
$\sigma^2 \simeq1.7 \, ({k_BT})^2$, however, was observed at a 
larger, optimal velocity $v^\ast \simeq 200$ nm/s,
with an error bar of \, $0.03$. \, For \, the \, unbiased 

\newpage

\vspace*{-6mm}

\noindent
comparison in Fig.\,\ref{1201_3a.fig},
we have chosen the optimal values $\omega^\ast$ and $v^\ast$ from 
the data shown in Fig.\,\ref{opti.fig}.

{\sl Discussion and summary -}
We have proposed a new method for recovering the free energy profile
$G(z)$ of biomolecules in single molecule pulling experiments by 
combining the new periodic ramp (\ref{cl}) with Jarzynski's
identity for recovering equilibrium properties from 
non-equilibrium data. The simulated data in Fig.\,\ref{1201_3a.fig}
show that the periodic ramp delivers significantly more
accurate data than the traditional linear loading (\ref{linear}),
under the constraint of equal experimental effort.
An additional advantage of the periodic ramp is the fact that 
the measuring time may be chosen as long as necessary to collect the 
sufficient amount of data to recover the barrier regions of the free 
energy profile, which are hard to determine by previous methods.

The driving frequency $\omega$ and preloading offset $x_0$ of the 
periodic ramp (\ref{cl}) provide handles to optimize its
performance (see Fig.\,\ref{opti.fig}). As our model case study
has shown, 
the frequency should be large enough to drive the
system out of equilibrium, but still smaller than the 
spontaneous transition rate under preloading. Whether the optimal
frequency  is always of this order remains to be seen in
more comprehensive systematic investigations left for future work.

Our method can be extended to probing free energy profiles
with different transitions between a number of metastable states. 
If these states are sufficiently separated, by several nanometers and 
energy barriers of several $k_B T$, each transition can be selected 
by using suitable values for preloading offset $x_0$ and amplitude 
$a$ in (\ref{cl}). Subsequently, these parts of the free energy 
profile may be fitted together by adjusting free additive constants 
on each part, similarly as shown in \cite{Jensen}.

\end{document}